\documentstyle[12pt]{article}
\begin{document}
\baselineskip6.8mm
\renewcommand{\theequation}{\thesection.\arabic{equation}}
\title{Supermembranes and Superstrings with
Extrinsic Curvature}
\author{\\
F. Aldabe${}^{*}$\,
and A.L. Larsen${}^{||}$}
\maketitle
\noindent
{ \em
Theoretical Physics Institute, Department of Physics, \
University of
Alberta, Edmonton, Canada T6G 2J1}
\vskip 12pt
\begin{abstract}
\baselineskip=1.5em
In a recent paper Townsend suggested to associate to the
D=11 solitonic membrane of N=1 supergravity a certain
thickness, and then to identify this membrane with the
fundamental supermembrane. By integrating out
the 8 transverse dimensions of the "thick" solitonic
membrane, we show that the resulting world-volume action
indeed contains all the usual supermembrane terms,
as well as background curvature terms and
extrinsic curvature terms, which are
believed to render the membrane spectrum discrete. We also
outline the analog derivation for the "thick" D=10 solitonic
string solution of N=IIA supergravity.  The resulting
world-sheet action contains the usual type IIA superstring
terms, as well as extra terms whose presence can be
interpreted as a rescaling of the background metric,
thus preserving kappa-symmetry and conformal invariance.

\end{abstract}
\noindent
\\
\noindent
$^{*}$Electronic address: faldabe@phys.ualberta.ca\\
$^{||}$Electronic address: alarsen@phys.ualberta.ca
\newpage
\section{Introduction}
\setcounter{equation}{0}
The supermembrane \cite{TBS} seems to play a fundamental
role in
the unification of string theories, which are then interpreted
as different perturbative expansions of the supermembrane.
However, a major drawback of the supermembrane is that its
spectrum is
continuous \cite{Wit}, and therefore it cannot reproduce the
particle
spectrum we observe.  The membrane has
a continuous
spectrum because no energy is required to create a
spike  of
arbitrary height and of zero area \cite{Wit}.
Despite this, string-membrane duality in 10 dimensions
between the type IIA superstring and the supermembrane
requires that
the type IIA fundamental string  be identified with the
solitonic string
of 10 dimensional N=IIA supergravity, and that the
supermembrane be
identified with the solitonic membrane of 11 dimensional N=1
supergravity \cite{T}.

As suggested in \cite{T}, this identification between
fundamental and
solitonic membrane can be used to yield a fundamental
membrane which has
a thickness.  Spikes on this membrane will then necessarily
have a non-zero
area, and therefore energy will be required to create such
spikes.  Thus,
a membrane with thickness is believed to have a discrete
spectrum with a graviton
multiplet as its ground state \cite{TBS2}.
But we also must identify the fundamental and solitonic
strings of the
type IIA theory.  Such an identification will lead to a
fundamental
 string  with a thickness.   Alternatively, double dimensional
reduction of the membrane \cite{DIHS}  with a thickness, will
also lead to
a fundamental
string with a thickness.  It then seems that string-membrane
duality along
with discreteness of the supermembrane spectrum
should lead to new fundamental string theories which contain
thickness terms.

Here we derive to leading order  the
thickness terms of the supermembrane  by integrating out
the transverse dimensions of the solitonic membrane of 11
dimensional
N=1 supergravity. The thickness terms consist of
background curvature terms as well as extrinsic curvature
terms, which are
believed to render the membrane spectrum discrete. We also
derive to leading order
the thickness terms for the type IIA superstring by
integrating
out the transverse dimensions of the solitonic string of 10
dimensional
N=IIA supergravity. We
show that the fundamental
string with thickness is equivalent to another string theory
without  thickness terms,  but in a different vacuum where
the background metric has acquired corrections of ${\cal
O}(\alpha').$
In conclusion, we argue that the leading order thickness
terms
obtained by identifying the fundamental membrane (string)
with the solitonic
membrane (string)  leads to a discrete spectrum for the
membrane, but the identification does not lead
to new string theories.

In Section 2 we review the solitonic membrane solution of 11
dimensional
N=1 supergravity.  In Section 3  we derive the "thick"
supermembrane from
the solitonic membrane and explain why the thickness terms
suppress
the spikes.  In Section 4 we derive the "thick" fundamental
string from the
solitonic string of the 10 dimensional
N=IIA supergravity theory, and show how the thickness
terms redefine the
vacuum.  Some concluding remarks are given in Section 5.

\section{The Solitonic Membrane of D=11 Supergravity}

The field equations in the bosonic sector of N=1
supergravity in 11 dimensions, which can be derived from
the action \cite{Nahm},
\begin{eqnarray}
S_{11}=\frac{1}{2\kappa^2}\int d^{11}Z
\;[\hspace*{-2mm}&\;&\hspace*{-7mm}\sqrt{-g}\;
(R-\frac{1}{48}
F_{MNOP}F^{MNOP})\nonumber\\
\hspace*{-8mm}&+&\hspace*{-2mm} \frac{1}{12^4}
\epsilon^{MNOPQRSTUVW}F_{MNOP}F_{QRST}A_{UVW}],
\end{eqnarray}
admit  a "solitonic membrane" solution given by:
\begin{equation}
ds^2=\left( 1+\frac{K}{r^6}\right) ^{-2/3}
\hspace*{-2mm}\eta_{\mu\nu} dx^\mu dx^\nu+
\left( 1+\frac{K}{r^6}\right) ^{1/3}\hspace*{-2mm}\delta_{nm}
dy^n dy^m,
\end{equation}
\begin{equation}
A_{012}=\pm\left( 1+\frac{K}{r^6}\right) ^{-1},
\end{equation}
as was shown by Duff and Stelle \cite{DS}.
Here $M, N=0,1,...,10;\;\;\mu, \nu=0,1,2;\;\;m, n=3,4,...10\;$
and $\;r^2\equiv\delta_{nm}y^n y^m.$ The membrane is
parametrized by the constant $K,$ which has dimension of
$(length)^{6},$ and which is taken to be positive.

Invariants constructed from $g_{MN}$ and $A_{MNP},$ such
as
$R,\;F^2,..$ are roughly stepfunctions. Indeed:
\begin{equation}
R\sim F^2\sim\frac{K^2}{(K+r^6)^{7/3}},\;\;\;\;\;\;\;r>0.
\end{equation}
Already at this stage, it is natural to associate to this
membrane a certain "thickness", namely the thickness of the
"tube" in which the invariants $R,\;F^2,..$ are significantly
different from zero. In this sense, the thickness of the
membrane (2.2)-(2.3),
would be a number somewhat less than $1$ in units of
$K^{1/6} .$
However, the solution (2.2)-(2.3) is only valid for $r>0.$ To
continue the solution to $r=0,$ Duff and Stelle introduced a
membrane-source
in the supergravity field equations \cite{DS}. This leads to a  
consistent
solution everywhere, but introduces (besides the source) a
$\delta$-function singularity in the scalar curvature at $r=0.$  
However, as
shown by Tseytlin \cite{tsey1,tsey2}, one may expect that  
$\alpha'$-corrections
will smooth out the singularity and thus give rise to a finite  
thickness of the
order of $\sqrt{\alpha'}.$
Duff, Gibbons and Townsend \cite{DGT}, on the other hand,
suggested to avoid the source by analytical continuation of
$r$ to imaginary values. They introduced a new radial
coordinate $\hat{r}\equiv(r^6+K)^{1/6}$ and analytically
continued it to $\hat{r}=0.$ In that case, the $\delta$-function
singularity at $r=0$ is exchanged by an ordinary
Schwarzschild-like event
horizon surrounding a physical singularity at $\hat{r}=0$
\cite{DGT}. Now there is no problem of continuing the
solution
(2.2)-(2.3) through $r=0;$ one merely has to express the
solution in a coordinate system which is well-behaved at
$r=0.$ In particular, the expressions (2.4), which in fact are
regular at $r=0$ anyway, can be continued through $r=0$
also.

In a recent paper \cite{T}, Townsend argued that one
should associate to this analytically continued solitonic
membrane a finite thickness of the order of the radius of the
event horizon, i.e. again a thickness of the order $K^{1/6}.$
He further suggested to identify this solitonic membrane with
the fundamental supermembrane, which thereby
acquires a finite thickness.

\section{Solitonic Membrane and Extrinsic Curvature}
\setcounter{equation}{0}
The configuration (2.2)-(2.3) describes a flat static membrane,
and it is the only explicitly known membrane solution of N=1
supergravity in D=11. In some sense it plays the same role as
the flat static cosmic membrane, $\phi\sim\tanh(z),$ in
ordinary $\phi^4$-theory with Mexican hat potential,
originally found by Zeldovich, Kobzarev and Okun
\cite{Zeld}. Inspired by this analogy and the work on
generic curved dynamical cosmic membranes, see for
instance \cite{Greg}, we will now be interested in the
dynamics of a general curved solitonic membrane solution of
the  N=1 supergravity field equations in D=11.

The idea is to insert the solution for a generic curved
solitonic membrane, which is of course not known explicitly
in terms of $g_{MN}$ and $A_{MNP},$ into the action (2.1),
and to integrate out the 8 transverse dimensions (in a first
approximation, one may think of this procedure as
considering small fluctuations around the flat static
membrane). This will lead to an effective action
determining the dynamics of the core of the solitonic
membrane, and it is eventually to be identified with the
world-volume action of the "thick" fundamental
supermembrane. The procedure used to obtain this effective
action
is well-known from the theory of
cosmic defects (
see for instance \cite{Greg}), and it can be straightforwardly
generalized
to p-branes in supergravity.
However,
in previous discussions of p-branes in supergravity, the
effective world-volume action has instead been postulated
directly from its symmetries (see for instance
\cite{Hug,HD,Cur,Lu}),  and
therefore does not contain any
knowledge about possible
thickness terms. In
the case of the solitonic membrane, where we do not want to
erase completely the effects of the finite thickness, it is more
appropriate to actually integrate out the transverse
dimensions explicitly.

In order to perform the integration over the transverse
dimensions, we will assume that the solitonic membrane
under consideration is relatively smooth and slowly varying,
in the sense that it locally looks like the flat static solution
(2.2)-(2.3).
It means that we will assume that all invariants constructed
from $g_{MN}$ and $A_{MNP}$ are constant on the core of
the membrane and that, close to the core, they only depend
on the
transverse coordinates. These assumptions are similar to the
assumptions usually made for the Higgs field in cosmic
membrane theory \cite{Greg}; in our case,
the role of the Higgs field is played by the
background fields $g_{MN}$ and $A_{MNP}$ themselves.

It is convenient to define coordinates on the core of the
membrane:
\begin{equation}
Z^M(x^\mu,y^m)=X^M(x^\mu)+y^m N^M_m(x^\mu),
\end{equation}
and to introduce Riemann normal coordinates:
\begin{equation}
g_{MN}=\eta_{MN}-\frac{1}{3}y^n y^m N^P_n N^Q_m
R_{MPNQ}(y=0)+{\cal O}(y^3).
\end{equation}
We use the same notation as in (2.2)-(2.3), i.e.
$M, N=0,1,...,10;\;\;\mu, \nu=0,1,2;\;\;m, n=3,4,...10.$

The normal vectors $N^M_m$ fulfill:
\begin{equation}
\eta_{MN}X^M_{,\mu} N^N_m=0,\;\;\;\;\;\;\;\;\eta_{MN}N^N_n
N^M_m=\delta_{nm},\label{nor}
\end{equation}
as well as the completeness relation:
\begin{equation}
\eta^{MN}=\gamma^{\mu\nu} X^M_{,\mu}
X^N_{,\nu}+\delta^{nm}
N^M_m N^N_n,
\end{equation}
where $\gamma_{\mu\nu}$ is the induced metric on the core
of the membrane:
\begin{equation}
\gamma_{\mu\nu}=\eta_{MN} X^M_{,\mu} X^N_{,\nu}.
\end{equation}
{}From (3.3)-(3.4) it can be shown that the normal vectors
$N^M_m$ are
functions of $X^M$
but not of $X^M_{,\mu},$ in the sense that the
transformation laws under
world-volume reparametrizations are given by:
\begin{equation}
\delta X^M=\xi^\mu X^M_{,\mu},\;\;\;\;\;\;\;\;\delta
N^M_m=\xi^\mu
N^M_{m,\mu}.
\end{equation}
We also define the extrinsic curvature $K_{\mu\nu n}$ and
torsion $\omega_{m n\nu}:$
\begin{equation}
K_{\mu\nu n}=\eta_{MN} X^M_{,\mu} N^N_{n,\nu},\;\;\;\;\;\;\;\;
\omega_{m n\nu}=\eta_{MN} N^M_m N^N_{n,\nu}.
\end{equation}
The idea is now to integrate out the 8 transverse
$y$-coordinates in the action (2.1). The simplest way to do
this, is to Taylor-expand the integrand in  powers of $y, $
and to perform the integration only for "small"  $y$ ($\sim$
one unit in the thickness of the membrane). This is a
reasonable approximation of the actual integral, provided the
integrand resembles a "narrow" stepfunction.
This is actually the case for the flat static membrane
(2.2)-(2.3), and
we can expect this to hold also for a more general smoothly
curved and slowly varying membrane.

Let us first consider the volume element in (2.1).
It is a standard exercise to compute the metric $g_{MN}$ in
the $(x,y)$-coordinates:
\begin{eqnarray}
g_{\mu\nu}=\gamma_{\mu\nu}\hspace*{-2mm}&+&
\hspace*{-2mm}2y^m K_{\mu\nu m}+
y^n y^m[
K_{\mu\rho
m}K_\nu\;^{\rho}\;_n+\omega_{pm\mu}\omega^p\;_{n\nu}
\nonumber\\
\hspace*{-2mm}&-&\hspace*{-2mm}\frac{1}{3}N^P_m
N^Q_n
X^M_{,\mu} X^N_{,\nu}R_{MPNQ}(y=0)]+{\cal O}(y^3),
\end{eqnarray}
\begin{equation}
g_{\mu m}=y^n\omega_{m n\mu}-\frac{1}{3}y^n y^p N^P_n
N^Q_p N^N_m
X^M_{,\mu}R_{MPNQ}(y=0)+{\cal O}(y^3),
\end{equation}
\begin{equation}
g_{m n}=\delta_{m n}-\frac{1}{3}y^p y^q N^P_p N^Q_q
N^M_m N^N_n R_{MPNQ}(y=0)+{\cal O}(y^3).
\end{equation}
It follows that:
\begin{eqnarray}
\sqrt{-g}=\sqrt{-\gamma}\;[1+y^n
K^\mu_{\;\;\mu
n}\hspace*{-2mm}&+&\hspace*{-2mm}\frac{1}{2}y^n
y^m(K^\mu\;_{\mu n}K^{\nu}\;_{\nu m}-K^{\mu\nu}\;_n
K_{\mu\nu m}\nonumber\\
\hspace*{-2mm}&-&\hspace*{-2mm}\frac{1}{3}N^M_m N^N_n
R_{MN})+{\cal O}(y^3)].
\end{eqnarray}

Next we consider the Lagrangian. As already explained, we
assume that all invariants $(R,\;F^2,...)$ constructed from
$(g_{MN},\;A_{MNP})$ are constant on the core of the
membrane, and close to the core, they only depend on the
$y$-coordinates.
It should be stressed that these assumptions are not made
to simplify the mathematics; they are merely conditions for
the actual existence of membrane-shaped solutions of the
D=11 supergravity field equations.
Thus, considering a membrane which locally
resembles the flat static membrane (2.2)-(2.3), we can
assume:
\begin{equation}
R=R(y=0)+{\cal O}(y^2),
\end{equation}
\begin{equation}
F^2=F^2(y=0)+{\cal O}(y^2).
\end{equation}
These expressions will be valid for small $y,$ that is, inside
the range of the $y$-integration, c.f. the comments after
equation (3.7). Notice also that the assumptions (3.12)-(3.13)
are relatively "modest" for a smooth and slowly varying
membrane. For the flat static membrane (2.2)-(2.3), the first
corrections to the constant values at $y=0$ are actually of
${\cal O}(y^6).$

Generally there is no reason for $g_{MN}$ and $A_{MNP}$ to
be functions of the $y$-coordinates only (although this is the
case for the flat static membrane (2.2)-(2.3)). Contrary to
$(R,\;F^2,...),$ they are not invariants and therefore will
depend on the specifically chosen coordinates and
three-form gauge. However, to fulfill equation (3.13),
$A_{MNP}$
must have the form:
\begin{equation}
A_{MNP}\; :\;\;\;\left\{ \begin{array}{lll}
A_{\mu\nu\rho}=A_{\mu\nu\rho}(X(x))+{\cal O}(y^3) \\
A_{\mu\nu m}=A_{\mu m n}={\cal O}(y^3) \\
A_{m n p}=y^q C_{m n p q} + {\cal O}(y^3) \end{array}\right.
\end{equation}
where $C_{m n p q}$ are constants, while $A_{\mu\nu\rho}$
are arbitrary functions of $X.$ It follows that:
\begin{equation}
\epsilon^{MNOPQRSTUVW}F_{MNPQ}F_{QRST}A_{UVW}
\sim
\epsilon^{\mu\nu\rho}X^M_{,\mu}X^N_{,\nu}X^P_{,\rho}
A_{MN
P}(y=0)+{\cal O}(y^2).
\end{equation}
Thus, the topological term of 11 dimensional N=1
supergravity is responsible for
the Wess-Zumino term of the supermembrane action.

Using also the following integrals:
\begin{equation}
\int^{'}d^8 y\sim K^{4/3},
\;\;\;\;\;\;\;\;\int^{'}y^n\;d^8 y=0,\;\;\;\;\;\;\;\;\int^{'}y^n  
y^m\;d^8
y\sim
K^{5/3}\delta^{nm},
\end{equation}
where the prime denotes integration only over the "thickness"
($\sim$ one unit in $K^{1/6}$), we get the following result
from (2.1) to zeroth order in $y$ (after a constant redefinition
of $A_{MNP}$ and $g_{MN}$):
\begin{equation}
S_{M}^{(0)}\sim\frac{K}{\kappa^2}\int d^3 x [
\sqrt{-\gamma}+
\epsilon^{\mu\nu\rho}X^M_{,\mu}X^N_{,\nu}X^P_{,\rho}
A_{MN
P}],
\end{equation}
that is, the usual bosonic sector of the supermembrane
action
\cite{TBS}.
In cosmic membrane theory, this would describe the
zero-thickness limit \cite{Greg}. However, for the solitonic
membrane it
has no physical meaning to take this limit because of the
singularity in the center of the core. In other words, we must
consider a membrane of finite thickness or equivalently of
finite $K$. This corresponds to
keeping terms of higher order in $y$ in the action (2.1). To
second order in $y,$ we get from (2.1):
\begin{equation}
S_{M}^{(2)}\sim\frac{K^{4/3}}{\kappa^2}\int d^3 x
\;\sqrt{-\gamma}\;  [K^{\mu}\;_{\mu
n}K_{\nu}\;^{\nu n}-
K^{\mu\nu}\;_n
K_{\mu\nu}\;^n-\frac{\delta^{nm}}{3}R_{MN}N^M_m
N^N_n],
\end{equation}
that is, the extrinsic curvature terms as well as a term due to
the curvature of the D=11 background. Notice that to this order in  
$y,$ the
corrections (3.18) come entirely from the Jacobian of the coordinate
transformation (3.1), that is to say, from equation (3.11). To higher  
orders
there will be corrections also from the Lagrangian, and in  
particular, there
will be corrections to the Wess-Zumino term.

It must be stressed that the action (3.17)-(3.18) has been
derived only for very special D=11 backgrounds, namely
backgrounds that admit membrane-shaped solutions to the
D=11 supergravity field equations. However, to identify
(3.17)-(3.18) with the action of a new "thick" fundamental
supermembrane, we extrapolate the result to
arbitrary $(g_{MN},\;A_{MNP}).$ It is important that the
extrinsic curvature terms in (3.18) survive, even in the
background of Minkowski space. Extrinsic curvature terms
have been discussed in
other areas of physics, and are generally known to suppress
"spikes", see for
instance \cite{Poly}, thus it is reasonable to believe that the
spectrum, as obtained (non-perturbatively) from
(3.17)-(3.18), will be discrete.
Notice also that in the background of Minkowski space, by
using the Gauss-Codazzi equation \cite{Eis}, the action (3.18)
reduces to:
\begin{equation}
S_{M}^{(2)}\sim\frac{K^{4/3}}{\kappa^2}\int d^3 x
\;\sqrt{-\gamma}\;\; ^{(3)}R(\gamma),\label{r3}
\end{equation}
where $\;^{(3)}R(\gamma)$ is the scalar curvature of the
membrane.  This means that even in flat Minkowski space a
thickness term is present to insure that the spectrum of the
supermembrane is discrete.

A question that must be addressed is the preservation of
supersymmetry
and kappa-symmetry after including the leading order
thickness
terms \cite{cur,gau}.  For simplicity we treat only the case in
which the target space is flat.  Although supersymmetry is
not apparent
in the action obtained from the solitonic membrane, it is a
standard procedure to construct
the supersymmetric version, see for instance \cite{TBS}.
The
expression (\ref{r3}) will then be invariant
under the supersymmetry transformations of \cite{TBS2},
which leave the leading
order supermembrane action (3.17) invariant.   As far as
kappa-symmetry is
concerned,  to leading order the action is kappa-invariant
\cite{TBS2}.
This means that to leading order the spectrum is
supersymmetric.  However,
the presence of (\ref{r3}) may lead to kappa-symmetry
breaking (unless some compensating terms can be added)
and therefore
to a supersymmetry breaking of the particle spectrum. We
hope to address this issue somewhere else; see also the
comments in the Conclusion.

\section{Solitonic String and the Rescaled Metric }
\setcounter{equation}{0}
The solitonic string solution in 10 dimensions \cite{HD} was
originally constructed from the N=1 supergravity field
equations. For our purposes, it is however convenient to
consider the full N=IIA supergravity action
in D=10:
\begin{eqnarray}
S_{10}=\frac{1}{2\kappa^2}\int d^{10}Z\hspace{-4mm}
&[&\hspace*{-2mm}
\sqrt{-g}
\;\;\{e^{-2\Phi}\;(R+4(\nabla\Phi)^2-\frac{1}{3}H^2)
-G^2-\frac{1}{12}\tilde{F}^2\}\nonumber\\
&-&\hspace*{-2mm}\frac{1}{288}
\epsilon^{MNOPQRSTUV}F_{MNOP}F_{QRST}B_{UV}],
\end{eqnarray}
from which the N=1 supergravity action is obtained by
truncation, and we consider only the bosonic part.
Here $\Phi$ is the dilaton, $G=dA,\;H=dB,\;F=dC$ and
$\tilde{F}=dC+2A\wedge H$ ($A$ is the one-form, $B$
the two-form and $C$ the three-form).
The solitonic string solution \cite{HD} is then given by:
\begin{equation}
A=0,\;\;\;\;C=0,
\end{equation}
\begin{equation}
ds^2=\left( 1+\frac{K}{r^6}\right) ^{-3/4}
\hspace*{-2mm}\eta_{\mu\nu} dx^\mu dx^\nu+
\left( 1+\frac{K}{r^6}\right) ^{1/4}\hspace*{-2mm}\delta_{nm}
dy^n dy^m,
\end{equation}
\begin{equation}
B_{01}=\pm \left( 1+\frac{K}{r^6}\right) ^{-1},\;\;\;\;\;\;\;\;
e^{\Phi}=\left( 1+\frac{K}{r^6}\right) ^{-1/2},
\end{equation}
and we use the similar notation as in (2.1)-(2.3):
$M, N=0,1,...,9;\;\;\mu, \nu=0,1;\;\;m, n=2,3,...9.\;$ Notice also
that the line
element (4.3) is expressed in Einstein frame:
\begin{equation}
g_{MN}^{String}=e^{\Phi/2}g_{MN}^{Ein.}
\end{equation}
As for the solitonic membrane, there is a $\delta$-function
singularity at $r=0,$ corresponding to the string-source.
However, by a construction following the steps of the
membrane-case, but involving also a rescaling of the metric
and a reinterpretation of the dilaton \cite{DGT}, the source
can be
removed by analytical continuation beyond $r=0.$ Again it is
found that $r=0$ becomes a mere coordinate singularity (a
horizon) surrounding a true curvature singularity in the
analytically continued metric. Thus the situation is very
similar to the case of the membrane. The solitonic string
(4.2)-(4.4) is the straight static string-shaped solution to the
field equations obtained from the action (4.1). Following the
derivation of Section 3, it is then straightforward to obtain
the effective world-sheet action describing the core of a
more general curved non-static string solution, so we just
present the results
here without going into the computational details.

The analog of (3.17) becomes:
\begin{equation}
S_{S}^{(0)}\sim\frac{K}{\kappa^2}\int d^2 x \;
[\sqrt{-\gamma}+
\epsilon^{\mu\nu}X^M_{,\mu}X^N_{,\nu}B_{MN}],
\end{equation}
that is, the usual bosonic sector of the superstring action.
Notice that the WZ-term was obtained from the $F\wedge
F\wedge B$-term in (4.1), by a decomposition of $B$ similar
to
(3.14). To second order in $y$ we get the terms:
\begin{equation}
S_{S}^{(2)}\sim\frac{K^{4/3}}{\kappa^2}\int d^2 x
\sqrt{-\gamma}\;[ W(X)+ \; ^{(2)}R(\gamma)],\label{r4}
\end{equation}
where:
\begin{equation}
W(X)=\frac{5}{3}\delta^{nm}R_{MN}N^M_m
N^N_n-\delta^{mp}\delta^{nq}R_{MNPQ}N^M_m N^N_n
N^P_p N^Q_q,
\end{equation}
and we have used the Gauss-Codazzi equation \cite{Eis}.

We should stress,
that as in the case of the membrane,
it makes no sense to take the zero thickness limit (and thus
get rid of (\ref{r4})) because of the singularity in the center of
the string core.  As in
the membrane-case, we now take the action (4.6)-(4.7) and
extrapolate it to arbitrary $(g_{MN},\;B_{MN}).$ In Minkowski
space we just get the usual type IIA superstring action
(bosonic sector), while the $W(X)$-term in (4.7) must be
taken into
account in a general curved background.

Let us consider (4.6)-(4.7) in a little more detail. The second
term in (4.7) is topological and therefore can be skipped,
because the action is two-dimensional. Furthermore,
introduce the
string tension:
\begin{equation}
\frac{K}{\kappa^2}\sim\frac{1}{\alpha'},\;\;\;\;\;\;\;\;
\kappa^2\sim(\alpha')^4
\end{equation}
and write the action (4.6)-(4.7) in "Polyakov" form:
\begin{eqnarray}
S_{S}&\sim&S_0+\alpha' S_1\nonumber\\
S_0&=&\frac{1}{\alpha'}\int d^2
x\;[\sqrt{-h}\;h^{\mu\nu}\eta_{MN} X^M_{,\mu}
X^N_{,\nu}
+ \epsilon^{\mu\nu}X^M_{,\mu}X^N_{,\nu}B_{MN}]\nonumber\\
S_1&=&\frac{1}{\alpha'}\int
d^2x\;\sqrt{-h}\;h^{\mu\nu}\eta_{MN} W(X)
X^M_{,\mu}  X^N_{,\nu}
\end{eqnarray}
where  $h_{\mu\nu}$ is the Lagrange multiplier field.

The term  $S_1$ simply
rescales the background metric
by terms of ${\cal O}(\alpha')$.
A term which seems to be missing is the dilaton term,
$\; ^{(2)}R(\gamma)\Phi.$  A
similar
situation was encountered in \cite{Dab}, where the
world-sheet action
without thickness terms
for the heterotic string was derived.  At first one would
expect the
conformal invariance to be broken.  However, the dilaton term
appears
by considering an effective world-sheet action which
preserves conformal
invariance.  This is equivalent to adding the dilaton term in
order to insure
that the $\beta$-functions vanish, thus restoring conformal
invariance.
Since the  term $S_1$ only rescales the metric, it is
always possible to
find a background dilaton which will insure that the
$\beta$-functions vanish.
In terms of the background dilaton, $\Phi_s$,
 which defines the solitonic string, we may define the new
background dilaton
which restores conformal invariance
\begin{equation}
e^{2\Phi_{new}}=e^{2\Phi_s} [1+ \alpha'  W(X)].
\end{equation}
The constraints imposed on the background antisymmetric
tensor to achieve kappa-symmetry
are thus compatible with the vanishing of the
$\beta$-functions
\cite{T3}.

\section{Concluding Remarks}
\setcounter{equation}{0}
The main obstacle to accept the supermembrane to describe
the observed
particle spectrum is that its spectrum is continuous.  We
have shown,
as suggested by Townsend and needed by
string-membrane duality,
that the identification  between the solitonic and fundamental
membrane
yields a supermembrane action with thickness terms, which
are believed to suppress the
creation of spikes of zero area, and therefore render the
spectrum discrete.
In addition, as required by the $E_7(Z)$ invariance of the
spectrum \cite{HT}
of the type IIA superstring and by string-membrane duality,
we must also identify
the solitonic string with the  fundamental string.  We have
shown that this
identification does not yield a new string theory, but rather
that it
leads to a redefinition of the vacuum, at least when including only  
the lowest
order thickness terms.

An open problem on which we hope to report elsewhere, is
the possible breaking of
kappa-invariance and thus of the supersymmetry of the
particle spectrum of the membrane,
due to thickness terms.  If kappa-invariance is broken, it is
only softly broken
in the sense that the kappa-symmetry breaking terms will be
of ${\cal O}(\alpha').$
This would  mean that the requirement of string-membrane
duality not only yields
a supermembrane with a discrete spectrum but also a
spectrum which has a
softly broken supersymmetry.  Thus, this might be the much
sought
soft supersymmetry-breaking mechanism in string-membrane
theory.

\section*{Acknowledgements}

\setcounter{equation}{0}
The
work of  A.L. Larsen  was supported by NSERC.  F. Aldabe would like  
to thank
B. Campbell for discussion and W. Israel for ecouragement and  
support. We also
thank A. Tseytlin for drawing our attention to Refs.  
\cite{tsey1,tsey2}.


\begin{thebibliography}{11}

\bibitem{TBS} E. Bergshoeff, E. Sezgin, and P.K. Townsend,
Phys. Lett. {\bf B198}
(1987) 75.

\bibitem{Wit} B. De Wit, M. Luscher and H. Nicolai,
Nucl.Phys. {\bf B320}
(1989)
135.

\bibitem{T} P.K. Townsend, Phys. Lett. {\bf B350} (1995) 184.


\bibitem{TBS2} E. Bergshoeff, E. Sezgin and P.K.
Townsend, Ann. Phys. {\bf 185}
(1988) 330.

\bibitem{DIHS}
M.J. Duff, P.S. Howe, T. Inami and K.S. Stelle, Phys. Lett.
{\bf B191} (1987)
70.

\bibitem{Nahm} W. Nahm, Nucl.Phys. {\bf  B135} (1978)
149;
E. Cremmer, B. Julia, J. Scherk, S. Ferrara, L. Girardello and
P. Van
Nieuwenhuizen, Nucl.Phys. {\bf B147} (1979) 105.

\bibitem{DS} M.J. Duff and K.S. Stelle, Phys. Lett. {\bf B253}
(1991) 113.

\bibitem{tsey1} A. Tseytlin, Phys. Lett. {\bf B363} (1995) 223.

\bibitem{tsey2}A. Tseytlin, "Selfduality of Born-Infeld Action and  
Dirichlet
3-Brane of the Type IIB Superstring Theory", Imperial-TP-95-96-26,
hep-th/9602064.

\bibitem{DGT} M.J. Duff, G.W. Gibbons and P.K. Townsend,
Phys.Lett. {\bf B332} (1994) 321.


\bibitem{Zeld}Y.B. Zeldovich, I.Y. Kobzarev and L.B. Okun,
Sov. Phys. JETP {\bf
40} (1975) 1.

\bibitem{Greg}D. Garfinkle and R. Gregory, Phys. Rev. {\bf
D41} (1990) 1889;
R. Gregory, D. Haws and D. Garfinkle, Phys. Rev. {\bf D42}
(1990) 343;
R. Gregory, Phys. Rev. {\bf D43} (1991) 520.

\bibitem{Hug}J. Hughes, J. Liu and J. Polchinski, Phys. Lett.
{\bf B180} (1986)
370.

\bibitem{HD} A. Dabholkar, G.W. Gibbons, J.A. Harvey and
F. Ruiz-Ruiz, Nucl.
Phys. {\bf B340} (1990) 33.

\bibitem{Cur}C.G. Callan, J.A. Harvey and A. Strominger,
Nucl. Phys. {\bf B367}
(1991) 60.

\bibitem{Lu}M.J. Duff and J.X. Lu, Phys. Rev. Lett. {\bf 66}
(1991) 1402.

\bibitem{Poly}H. Kleinert, Phys. Lett. {\bf B174} (1986) 335;
A.M. Polyakov, Nucl. Phys. {\bf B268} (1986) 406;
H. Koibuchi, Phys. Lett. {\bf B242} (1990) 371.

\bibitem{Eis}L.P. Eisenhart, {\it Riemannian Geometry}
(Princeton University
Press, fifth printing, 1964).

\bibitem{cur}T. Curtright and P. van Nieuwenhuizen, Nucl Phys. {\bf  
B294}
(1987) 125.

\bibitem{gau}J.P. Gauntlett, K. Itoh and P.K. Townsend, Phys. Lett.  
{\bf B238}
(1990) 65.


\bibitem{Dab} A. Dabholkar, Phys. Lett. {\bf B357} (1995)
307.


\bibitem{T3} M.T. Grisaru, P. Howe, L. Mezincescu,
B.E.W. Nilsson and
P.K. Townsend, Phys. Lett. {\bf B162} (1985) 116.

\bibitem{HT} C.M. Hull and P.K.Townsend, Nucl. Phys. {\bf
B438} (1995) 109.







\end{thebibliography}
\end{document}